\documentclass{article}
\usepackage{spconf,amsmath,amssymb,graphicx,hyperref,booktabs,multirow,array}
\usepackage{cite}
\usepackage{newtxtext}

\title{Towards AI-Generated Music Plagiarism Detection as a Version Identification Problem}

\name{\shortstack{%
  Fotis Koutsikos$^{1}$ 
  Ioannis Prokopiou$^{2, 3}$
  Spyridon Kantarelis$^{1}$ 
  Vassilis Lyberatos$^{1}$\\
  Pantelis Vikatos$^{3}$ 
  Athanasios Aidinis$^{3}$ 
  Themos Stafylakis$^{2,4}$
  Athanasios Voulodimos$^{1}$
  Giorgos Stamou$^{1}$
  }
  }
\address{%
  $^1$National Technical University of Athens, Athens, Greece \\
  $^2$Athens University of Economics and Business, Athens, Greece \\
  $^3$Orfium, Athens, Greece \quad
  $^4$Archimedes/Athena R.C., Athens, Greece \\
  \fontsize{8}{2}\texttt{fkoutsikos12@gmail.com} \quad
  \fontsize{8}{2}\texttt{gian.prokopiou@aueb.gr} \quad
\fontsize{8}{2}\texttt{spyroskanta@ails.ece.ntua.gr} \quad
\fontsize{8}{2}\texttt{vaslyb@ails.ece.ntua.gr}
  }

\begin{document}
\small
%
\maketitle
\begin{abstract}
The rapid expansion of text-to-music generative models challenges traditional paradigms of music creation and intellectual property. Plagiarism in this context is rarely an absolute mathematical binary, but an ambiguous threshold negotiated over harmonic structure, melodic contours, or overall perceived stylistic character. In this work, we test the transferability of state-of-the-art music version identification architectures from the human-to-human cover domain to the human-to-AI plagiarism setting. To evaluate this task, we introduce \textsl{COPYCAT}, a benchmark derived from real-world plagiarism cases and extended through generative re-synthesis and digital signal processing obfuscations, yielding 350,654 evaluation pairs. We show that scalar distance thresholding collapses under generative re-synthesis, while a supervised framework leveraging coordinate-wise embedding shifts recovers the dispersed plagiarism signal, raising overall $F_{0.5}$ from $0.612$ to $0.803$.
\end{abstract}
\section{Introduction}

The rapid development of commercial and open-source text-to-music generative architectures, including \textsc{Suno}~\footnote{\url{https://suno.com/}}, \textsc{Udio}~\footnote{\url{https://www.udio.com/}}, and \textsc{MusicGen} \cite{musicgen}, has enabled high-fidelity audio synthesis from textual prompts and reference audio \cite{comanducci2024fakemusiccaps}. This progress has raised concerns regarding copyright infringement and potential plagiarism in AI-generated content \cite{batlleroca2024mira}. Unlike synthetic-audio detection, which determines whether content is generated, and source attribution, which seeks to identify training origins, plagiarism detection requires assessing whether generated audio retains high similarities to existing music despite substantial acoustic transformations \cite{smp}.

This problem is challenging both legally and computationally. On the legal side, infringement is assessed through specific elements - melody, harmony, rhythm - and overall impression, with no agreed standard of substantial similarity \cite{park2024quantitative}. At the same time, the scarcity of legally verified plagiarism cases limits the development and evaluation of automated methods \cite{saner2025music}. Meanwhile, the use of licensed music in generative AI motivates fine-grained similarity methods linking generated content to source recordings.

To date, computational approaches have primarily addressed human-to-human musical relationships \cite{yin2022originality}. Symbolic methods \cite{liu2023mesmf, park2022siamese} have been complemented by audio-based approaches such as the Music Segment Transcription System~\cite{mpd}, Comparative Imagery~\cite{saner2025music}, and MelodySim~\cite{melodysim}. Music version identification (MVI) offers a closely related task, recognizing corresponding recordings across changes in performance, instrumentation, tempo, and production \cite{doras2019coverdetectionusingdominant}. Whether such representations transfer to human-to-AI plagiarism remains underexplored.

This gap is particularly acute for generative re-synthesis \cite{crosvila2025aiarms, afchar2025aigenmusic}: unlike human covers, generative models introduce opaque acoustic distortions while preserving musical content, rendering surface-level similarity measures ineffective \cite{carlini2023extracting, de2026much}.


\begin{figure}[!t]
    \centering    
    \includegraphics[width=.64\columnwidth]{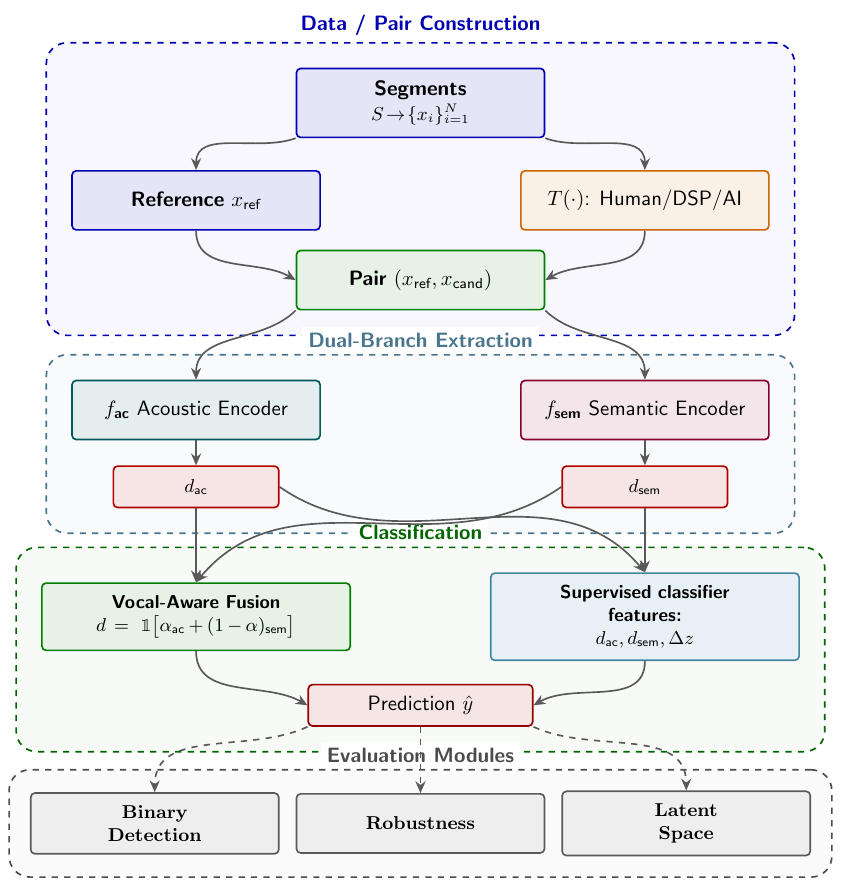}
    \caption{\small Overview of the proposed framework.}
    \label{fig:pipeline}
\end{figure}


Framing AI-generated plagiarism as a version identification task, in contrast to prior direct-similarity approaches, we introduce \textsl{COPYCAT} (Corpus Of Plagiarized Yet Camouflaged Artificial Tracks), a benchmark of 350,654 controlled pairwise comparisons constructed from the \textsl{SMP} dataset~\cite{smp} and expanded via three generative architectures combined with signal-level obfuscations. We evaluate two state-of-the-art MVI models, \textsc{CLEWS}\cite{clews} and \textsc{WEALY}\cite{wealy}, showing that scalar distances fail to fully exploit their representations under AI re-synthesis, whereas a supervised classifier over coordinate-wise embedding shifts recovers the dispersed plagiarism signal.

\section{Methodology}
\label{sec:meth}

\subsection{Problem Formulation}
\label{subsec:problem}

We formulate plagiarism detection as a pairwise similarity problem grounded in MVI: given an original and a candidate track, we assess whether the candidate can be regarded as a version or derivative of the original, regardless of whether it was produced by a human or a generative model. This formulation shifts the question from whether content is AI-generated to whether it is derivative, and enables us to investigate whether pretrained MVI representations transfer to the human-to-AI setting.

Formally, let $f_\theta$ denote an encoder mapping an audio segment $x$ to a latent embedding $z = f_\theta(x) \in \mathbb{R}^\ell$, and let $d(\cdot,\cdot)$ be a distance function over this embedding space. Given a candidate-original pair $(x_{\text{cand}}, x_{\text{ori}})$ with embeddings $z_{\text{cand}}, z_{\text{ori}}$, the plagiarism decision reduces to a threshold rule:
\[
\hat{y} =
\begin{cases}
1 \;\; (\text{plagiarism}), & \text{if } d(z_{\text{ori}}, z_{\text{cand}}) \leq \tau, \\[2pt]
0 \;\; (\text{otherwise}), & \text{if } d(z_{\text{ori}}, z_{\text{cand}}) > \tau.
\end{cases}
\]
Under this formulation, pairwise plagiarism detection reduces to the joint calibration of a distance function $d$ and a decision threshold $\tau$ in the embedding space. More generally, $\hat{y}$ can be produced by a supervised classifier operating on a richer pairwise representation (Section~\ref{subsec:modeling}). In either case, we treat $\hat{y}$ as a technical similarity judgment rather than a legal determination of substantial similarity.

\subsection{Benchmark Design}
\label{subsec:benchmark_design}

Evaluating this formulation requires a benchmark that covers both authentic human plagiarism and AI-generated derivatives with increasing levels of concealment. Since plagiarism typically occurs within specific passages rather than entire recordings, whole-track comparison would dilute the evidence with unrelated content. Therefore, our procedure begins by identifying the segments where similarity is alleged and extracting fixed-length audio excerpts. These excerpts serve as atomic anchors for all subsequent comparisons. Each anchor is then processed using a controlled set of transformations that represent progressively stronger forms of concealment: authentic human reinterpretation, deterministic digital signal processing (DSP) obfuscation applied to both originals and human covers, generative AI re-synthesis using multiple conditioning schemes, and combined AI and DSP transformations. Each transformed excerpt is paired with its anchor to form a positive example. Since randomly selected negatives would be too easy to distinguish, they can be organized into different difficulty levels, for example by mining hard distractors from the embedding space of a reference model. The resulting benchmark is organized around clear transformation categories, allowing performance to be evaluated as a function of concealment severity.

\subsection{Modeling}
\label{subsec:modeling}

Given a benchmark of such controlled pairs, we require a framework that can quantify the similarity between the two segments and render a plagiarism decision under the rule of Section~\ref{subsec:problem}. We begin by evaluating two pretrained MVI models as complementary branches: an acoustic branch (\textsc{CLEWS}) capturing timbral and structural similarity, and a semantic branch (\textsc{WEALY}) capturing lyrical similarity when vocals are present. We explore three progressively richer ways to combine their outputs.

\textbf{Threshold baseline:}
Following the decision rule introduced in Section~\ref{subsec:problem}, each branch is instantiated by selecting a distance metric $d$ and calibrating a threshold $\tau$ over the training partition. Both are jointly optimized to maximize $F_{0.5}$, which weights precision twice as heavily as recall, reflecting the high operational cost of false plagiarism accusations. Candidate distance metrics are compared exhaustively, and the pair $(d, \tau)$ that yields the highest $F_{0.5}$ on held-out folds is retained per branch.

\textbf{Score-level fusion:}
As a first attempt to exploit complementarity between the two branches, we perform vocal activity detection (VAD) and define a vocal-aware fusion rule over normalized scores ($\hat{d}$):
\[
d_{\text{fusion}} =
\begin{cases}
\alpha \, \hat{d}_{ac} + (1-\alpha) \, \hat{d}_{sem} & \text{if } V_{\text{pair}}=1, \\
\hat{d}_{ac} & \text{if } V_{\text{pair}}=0,
\end{cases}
\]
where $V_{\text{pair}}$ indicates vocal validity in both segments and $\alpha$ is calibrated on the training split to maximize $F_{0.5}$.

\textbf{Supervised shift modeling:}
Both preceding approaches collapse each embedding pair into a single scalar distance, potentially discarding coordinate-wise information relevant to plagiarism. We therefore additionally consider the coordinate-wise absolute shift vector
\[
\Delta z \;=\; \bigl(\,|z^{\text{cand}}_1 - z^{\text{ori}}_1|,\; \ldots,\; |z^{\text{cand}}_\ell - z^{\text{ori}}_\ell|\,\bigr) \;\in\; \mathbb{R}^\ell,
\]
and train a supervised classifier directly on it, rather than on signed shifts, as plagiarism is direction-agnostic. We evaluate three feature settings of increasing expressiveness: (1)~engineered features combining the branch-wise distances, statistical descriptors of $\Delta z$ (mean, standard deviation, percentiles, $L_2$-norm), stability-based statistics over the most and least volatile embedding dimensions, and vocal-validity flags derived from the VAD gating; (2)~the full $\Delta z$ vector; and (3)~a hybrid representation combining the engineered features with the most volatile embedding dimensions. This progression tests whether plagiarism evidence is concentrated in a few summary statistics or distributed across the latent space. In the hybrid setting, varying the dimensionality $K$ of the retained 
coordinates identifies the minimal subspace that preserves the plagiarism signal at reduced computational cost (see Section~\ref{subsec:results}). 

\section{Experiments}

\subsection{Experimental Setup}


\noindent\textbf{Dataset Construction:}\label{subsub:dataset} The \textsl{COPYCAT} dataset is built from \textsl{SMP}~\cite{smp}, to our knowledge the only publicly available collection pairing 70 real-world plagiarism allegations with segment-level audio timestamps across a broad stylistic spectrum. Using the timestamp annotations provided by \textsl{SMP}, we extract 20-second mono segments at 22,050\,Hz around the alleged regions of similarity, yielding 1,219 anchor segments. Each anchor is decomposed into vocals, bass, drums, and other stems via Hybrid Transformer Demucs v4~\cite{demucs}, enabling controlled vocal re-integration and stem-conditioned generation.

To populate the AI-derivative categories, we use three generative architectures exposing audio-conditioning interfaces, which commercial systems like \textsc{Suno} and \textsc{Udio} currently lack. \textsc{MusicGen} \cite{musicgen} (melody variant, 32\,kHz) conditions on chromagram features extracted from the anchor and generates a full re-synthesis. \textsc{AudioLDM~2} \cite{audioldm2} (audioldm\_48k variant, 48\,kHz) instead performs audio-guided diffusion generation, using the anchor as an acoustic reference. \textsc{MGE-LDM} \cite{mgeldm} occupies a distinct role: rather than regenerating the entire track, it preserves one anchor stem (bass, drums, or melodic/harmonic instruments) and inpaints the surrounding accompaniment, producing three stem-specific variants per anchor. This setting isolates the contribution of individual musical components to the plagiarism signal. Together, the three architectures yield 5,472 AI-derivative pairs.\looseness-1

Generations combine per-sample randomized text prompts, deliberately unrelated to the source style, so retained similarity stems from audio conditioning alone, with re-integrated human vocals under pitch and tempo augmentation. On top of these AI derivatives and the \textsl{SMP} originals, ten deterministic DSP transformations combining pitch shifts ($\pm2,\pm4$ semitones) and tempo modifications ($0.90$-$1.10\times$) expand the collection to 73,590 audio variants. Vocal validity is determined by an energy-based VAD filter on the anchor's Demucs-separated vocals \cite{de2026much}; since these are re-integrated into all derivatives, vocal presence is inherited, avoiding instrumental bleed in the mixed signal. Negative pairs are then mined in the \textsc{CLEWS} embedding space at three hardness levels-random, intra-category nearest (where categories correspond to distinct source $\times$ DSP combinations), and global nearest-with $\sim$79,000 pairs each. The final benchmark contains 350,654 pairwise comparisons (113,690 positive, 236,964 negative), with the positive-pair composition across the five transformation categories summarized in Table~\ref{tab:supervised_broad}. Since the \textsl{SMP} audio is commercially licensed, we release all reproducibility artifacts-including manifests, prompts, scripts, and source code-in the GitHub repository\footnote{\url{https://github.com/fotiskoutsikos/Plagiarism-Detection-System.git}}, rather than the raw audio.

\vspace{0.25\baselineskip}
\noindent\textbf{Model Configuration:}\label{subsub:modeling} The two MVI branches of Section~\ref{subsec:modeling} are instantiated with frozen pretrained models. For the acoustic branch we adopt \textsc{CLEWS} (Discogs-VI checkpoint, 1024-dimensional CQT embeddings), selected over \textsl{SHS100k} for its more realistic set-size distribution and state-of-the-art performance over prior CSI/MVI architectures (ByteCover \cite{bytecover3}, CoverHunter \cite{coverHunter}) \cite{clews}. The semantic branch relies on \textsc{WEALY}, extracting 512-dimensional embeddings from a frozen Whisper-Turbo backbone; it contributes only when both segments satisfy the vocal-validity condition of Section~\ref{subsub:dataset}.

On top of these two branches, the threshold baselines are calibrated independently, via stratified 5-fold cross-validation over four candidate distance metrics (euclidean, manhattan, cosine, pearson); the metric-threshold pair maximizing $F_{0.5}$ is retained per branch. The vocal-aware fusion weight $\alpha$ is selected by exhaustive grid search under the same $F_{0.5}$ criterion, so that all threshold-based configurations share a consistent calibration protocol.

The supervised classifier extends this setup along the coordinate-wise shift direction. A global \textsc{XGBoost} model is trained on each of the three feature settings introduced in Section~\ref{subsec:modeling}-engineered features, the full $\Delta z$ vector, and the hybrid representation-under stratified group 5-fold cross-validation grouped by original-track filename to prevent source leakage. Results are averaged over 10 seeds with 95\% confidence intervals. For the hybrid setting, the number $K$ of retained volatile \textsc{CLEWS} dimensions is chosen via a convergence sweep from 10 to 1024, identifying the point where predictive performance saturates relative to training cost (see Section~\ref{subsec:results}), with volatility ranked independently within each training fold to prevent leakage.\looseness-1

Beyond binary classification, we examine the embedding space through anchor-centered UMAP projections and distance-vs-DSP trendlines to assess how each source and transformation displaces derivatives in the latent space.

\subsection{Results}
\label{subsec:results}

\begin{table*}[!ht]
\small
\centering
\renewcommand{\arraystretch}{1}
\setlength{\tabcolsep}{4pt}
\resizebox{\textwidth}{!}{%
\begin{tabular}{l | cc | ccc | ccc | ccc}
\toprule
& \multicolumn{2}{c}{\textbf{\textsl{COPYCAT}}}
& \multicolumn{3}{c}{\textbf{\textsc{WEALY}}}
& \multicolumn{3}{c}{\textbf{\textsc{CLEWS}}}
& \multicolumn{3}{c}{\textbf{\textsc{XGBoost Hybrid Top-512}}} \\
\cmidrule(lr){2-3}  \cmidrule(lr){4-6} \cmidrule(lr){7-9} \cmidrule(lr){10-12}
\textbf{Category} & \textbf{Pairs} & \textbf{\%}
& \textbf{Prec.} & \textbf{Rec.} & \textbf{F$_{0.5}$}
& \textbf{Prec.} & \textbf{Rec.} & \textbf{F$_{0.5}$}
& \textbf{Prec.} & \textbf{Rec.} & \textbf{F$_{0.5}$} \\
\midrule
Human Plagiarism  & 3,757 & 3.3
& 74.2\% & 25.9\% & 54.0\%
& 98.0\% & 30.6\% & 68.0\%
& 97.3\% $\pm$ 0.1\% & 86.5\% $\pm$ 0.5\% & \textbf{94.9\% $\pm$ 0.2\%} \\

Human Plagiarism + DSP & 37,570 & 33.0
& 72.3\% & 24.8\% & 52.3\%
& 97.7\% & 27.4\% & 64.5\%
& 97.2\% $\pm$ 0.1\% & 83.1\% $\pm$ 0.4\% & \textbf{94.0\% $\pm$ 0.2\%} \\

Original + DSP & 12,190 & 10.7
& 51.8\% & 80.1\% & 55.7\%
& 96.9\% & 95.2\% & \textbf{96.5\%}
& 87.2\% $\pm$ 0.4\% & 99.3\% $\pm$ 0.0\% & 89.4\% $\pm$ 0.3\% \\

AI Generation  & 5,472 & 4.8
& 42.4\% & 51.5\% & 43.9\%
& 84.3\% & 15.0\% & 43.8\%
& 75.1\% $\pm$ 0.4\% & 56.5\% $\pm$ 0.3\% & \textbf{70.5\% $\pm$ 0.3\%} \\

AI + DSP & 54,701 & 48.1
& 36.4\% & 34.3\% & 36.0\%
& 79.0\% & 10.3\% & 33.9\%
& 72.0\% $\pm$ 0.5\% & 50.0\% $\pm$ 0.3\% & \textbf{66.2\% $\pm$ 0.4\%} \\

\midrule
Overall & 113,690 & 100.0
& 45.6\% & 36.6\% & 43.5\%
& 92.8\% & 26.0\% & 61.2\%
& 84.2\% $\pm$ 0.4\%
& 67.7\% $\pm$ 0.3\%
& \textbf{80.3\% $\pm$ 0.3\%} \\
\bottomrule
\end{tabular}%
}
\caption{\small Category-wise breakdown of the \textsc{WEALY}, \textsc{CLEWS} threshold baselines and the \textsc{XGBoost Hybrid Top-512} classifier, each fitted globally and evaluated per category. \textsc{XGBoost} results are reported as mean $\pm$ 95\% CI over 10 random seeds. Best F$_{0.5}$ per category in bold.}
\label{tab:supervised_broad}
\end{table*}







\vspace{0.25\baselineskip}
\noindent\textbf{Performance:}\label{subsub:performance} Among the four distance metrics, Manhattan distance yields the highest $F_{0.5}$ for \textsc{CLEWS}, while Cosine distance is optimal for \textsc{WEALY}. However, the semantic branch underperforms the acoustic one by a wide margin. Score-level fusion reflects this asymmetry: the grid search selects $\alpha=1.0$, effectively discarding the semantic signal. This outcome follows from the strong precision penalty of the $F_{0.5}$ criterion: any positive weight on the lower-precision semantic branch reduces overall precision faster than it recovers recall. These negative results motivate the coordinate-wise shift representation, which allows a supervised classifier to exploit fine-grained information from both branches that scalar distances discard.

The ablation described in Section~\ref{subsec:modeling} unfolds as follows. The vocal-validity flags are first identified as a shortcut: a classifier trained on the five flags alone reaches $F_{0.5}=0.836$, nearly matching the $F_{0.5}=0.853$ obtained when they are added to the engineered set. Since vocal-validity profiles correlate with transformation type, these flags let the classifier infer the category without examining musical content; they are therefore excluded. The remaining 24 engineered features yield only a marginal improvement over the \textsc{CLEWS} threshold ($F_{0.5}=0.617$ vs $0.612$), whereas the full 1024-dimensional \textsc{CLEWS} shift raises $F_{0.5}$ to $0.811$-indicating that the discriminative signal lies primarily in the raw coordinates rather than in scalar summaries. A convergence sweep over $K$ identifies $K=512$ ($F_{0.5}=0.792$ at 54\,s/fold) as the saturation point, within $0.02$ of $K=1024$ ($0.811$ at 119\,s/fold) and well above $K=256$ ($0.749$), leading to the final \textsc{Hybrid Top-512} model that combines the 24 engineered features with the 512 most volatile \textsc{CLEWS} dimensions and achieves an overall $F_{0.5}=0.803$. 

Feature importance analysis reveals a role for the semantic branch that scalar fusion could not expose: among engineered features alone, \textsc{WEALY} distances form the highest-ranked family, ahead of \textsc{CLEWS} distances. Individually, no raw dimension rivals these summaries, yet their cumulative effect drives the $+0.19$ $F_{0.5}$ gain over the engineered-only baseline.


As shown in Table~\ref{tab:supervised_broad}, the Hybrid model maintains high precision across all categories (from 97.3\% on human plagiarism down to 72.0\% on AI+DSP) while dramatically increasing recall: from 30.6\% to 86.5\% on human plagiarism and from 10.3\% to 50.0\% on AI+DSP-a near five-fold gain in the hardest category. The baseline retains an edge only on Original+DSP, where precision drops from 96.9\% to 87.2\%; the threshold's conservative behavior is inherently well-suited to trivial acoustic perturbations of an unmodified original.

\begin{figure}[!b]
    \centering
    \includegraphics[width=.85\columnwidth]{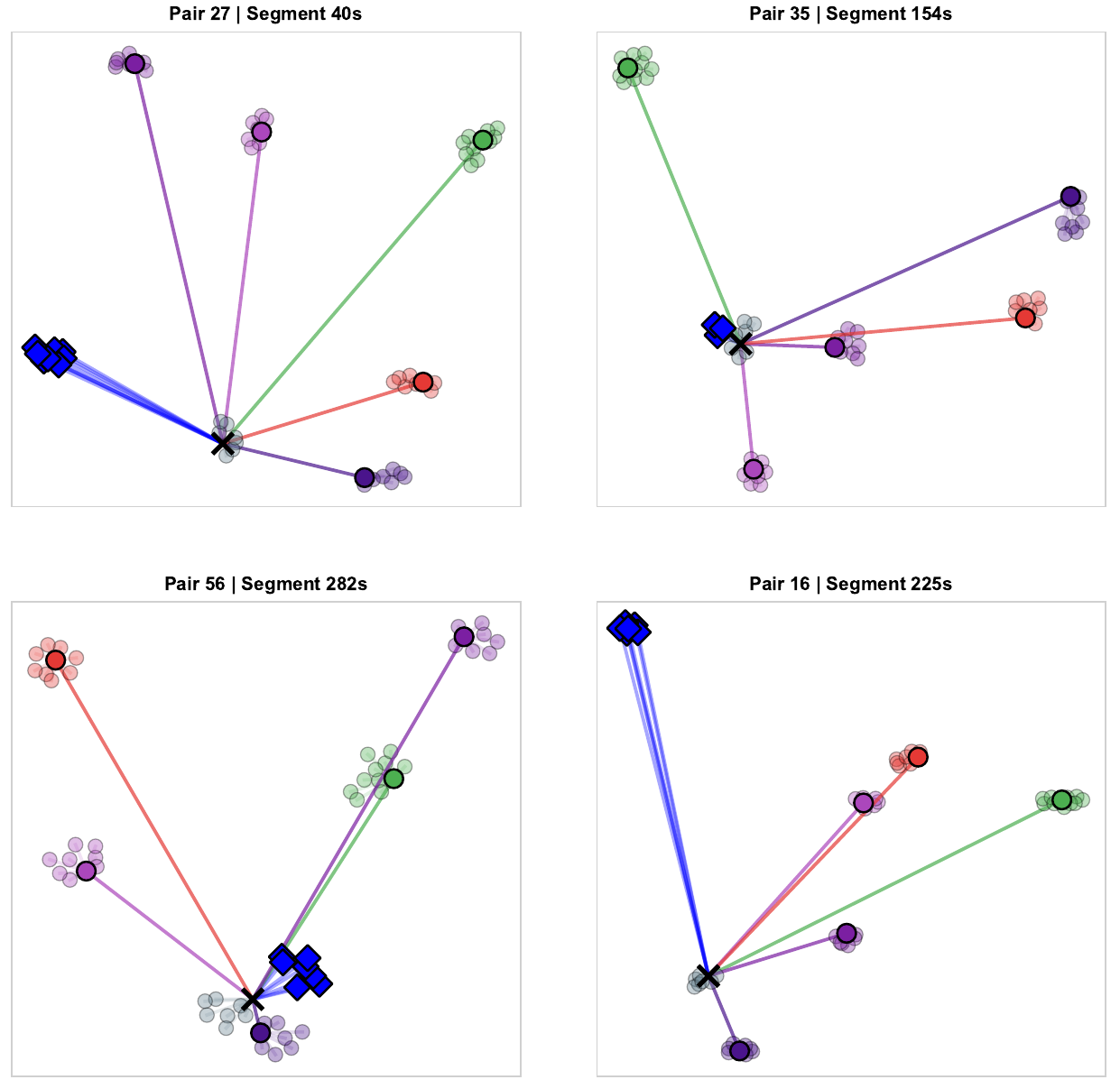}
    \caption{\small Origin-centered UMAP projection of \textsc{CLEWS} embeddings by transformation category (colors as in Fig.~\ref{fig:robustness}).}
    \label{fig:umap}
\end{figure}

\vspace{0.25\baselineskip}
\noindent\textbf{Latent Space:}\label{subsub:latent_space} To understand why the supervised shift model succeeds where scalar thresholding fails, we project the \textsc{CLEWS} embeddings into two dimensions via UMAP, centering each anchor at the origin so that derivative displacements directly reflect latent-space drift (Figure~\ref{fig:umap}). The projection reveals a source-specific geometric stratification: \textsc{AudioLDM 2} and \textsc{MusicGen} derivatives occupy the most distant neighborhoods, \textsc{MGE-LDM} variants cluster at intermediate radii depending on the preserved stem, and human covers lie closest to the anchor. Within each neighborhood, DSP-modified variants form peripheral shells around their base derivative, producing radial patterns that grow with concealment severity. This structure confirms that plagiarism-relevant information persists in a geometrically organized form-even for the hardest AI-derived cases that scalar distances place beyond the detection threshold (this stratification is consistent across the corpus and mirrored in the \textsc{WEALY} space).

\begin{figure}[!t]
    \centering
    \includegraphics[width=.95\columnwidth]{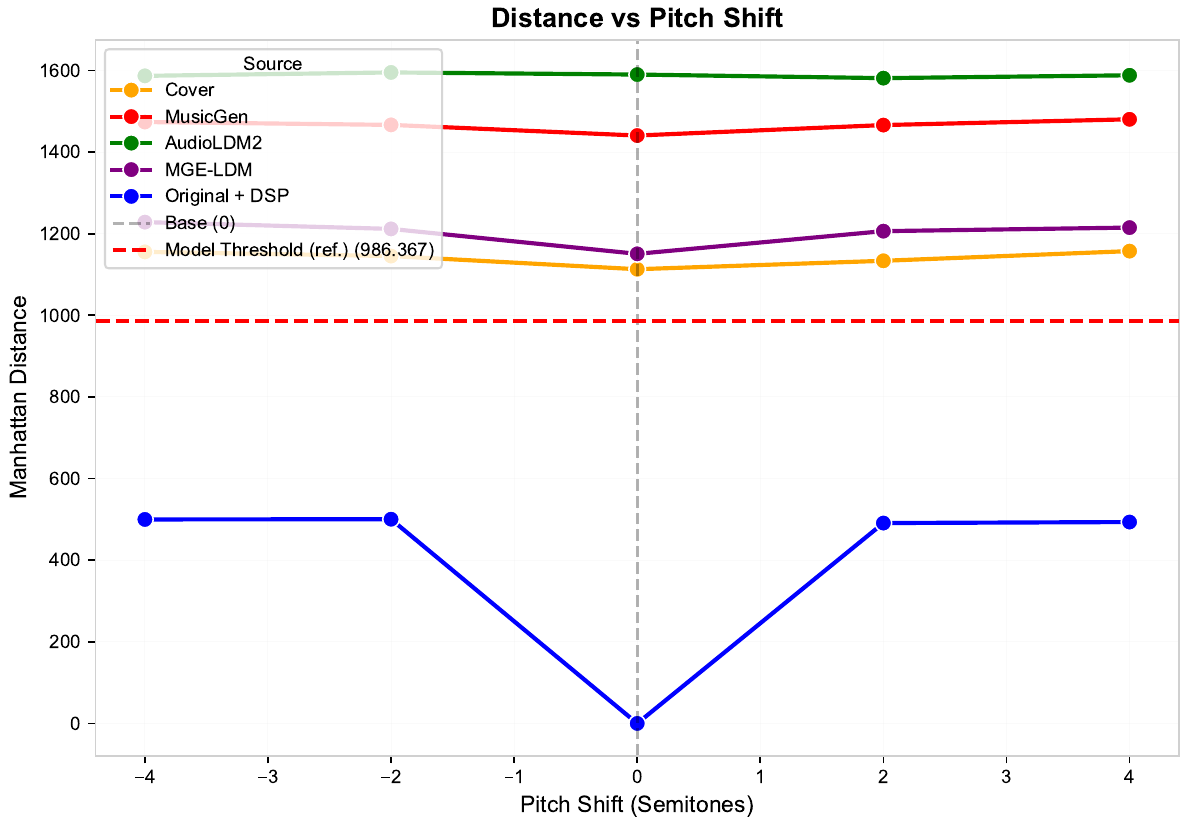}
    \caption{\small \textsc{CLEWS} pairwise distances. 
    Dashed line: calibrated threshold $\tau=986.4$.}
    \label{fig:robustness}
\end{figure}

\vspace{0.25\baselineskip}
\noindent\textbf{Robustness:}\label{subsub:robustness} Complementing the geometric view, we track raw \textsc{CLEWS} pairwise distances as a function of DSP intensity across all source categories (Figure~\ref{fig:robustness}; the calibrated threshold is overlaid as a visual reference only, and all measurements are threshold-independent). The trendlines reveal a four-tier stratification: pitch-shifted originals remain well within the detectable region across all intensities (tempo shifts behave analogously), human covers and \textsc{MGE-LDM} derivatives operate near the threshold with only marginal DSP sensitivity, and \textsc{MusicGen} and \textsc{AudioLDM 2} exhibit essentially flat trendlines far above the threshold. This saturation indicates that generative re-synthesis has already displaced the embeddings to near-maximal acoustic distance before any DSP is applied, leaving no residual sensitivity for a magnitude-based threshold to exploit. The discriminative signal in these hardest cases is therefore encoded not in the magnitude of the displacement but in its coordinate-wise structure - capturing not only how far a derivative drifts from its anchor but also in which directions - precisely the information that the Hybrid Top-512 classifier recovers through the raw $\Delta z$ representation. \textsc{WEALY} exhibits an analogous but complementary pattern in the semantic space, with pitch-sensitive rather than tempo-sensitive degradation, confirming that the two branches respond to distinct failure modes under adversarial DSP.

\section{Conclusions}

We reframed AI-generated music plagiarism detection as a music version identification problem between human originals and generative derivatives, evaluating pretrained MVI models on \textsl{COPYCAT}, a benchmark spanning human plagiarism, DSP obfuscation, and AI re-synthesis. Scalar distance thresholds over pretrained MVI embeddings fail on AI-derived cases not because the plagiarism signal is lost, but because it redistributes across embedding coordinates in ways that scalar distances cannot capture. Our \textsc{Hybrid Top-512} classifier recovers this signal from the coordinate-wise shift $\Delta z$, raising $F_{0.5}$ from 0.612 to 0.803, with the largest gains on AI+DSP. Detection difficulty scales with concealment: human plagiarism and simple DSP remain within reach, but generative re-synthesis combined with adversarial DSP pushes recall to its limits, confirming AI-mediated derivation as a fundamentally harder regime than human-to-human plagiarism. Given these limitations, the framework is best positioned as a triage mechanism that surfaces candidate cases for expert review, not as a substitute for legal judgment. While the semantic (\textsc{WEALY}) branch proved ineffective under scalar fusion, feature importance analysis confirmed its non-negligible contribution within the hybrid classifier. Future work will extend the benchmark to additional generative architectures, broader DSP perturbations, and lyric-preserving datasets~\cite{sonics}, evaluate other recent MVI systems \cite{araz_unified_2026} as alternative backbones, investigate fine-tuning of the MVI encoders, and address source attribution as a natural next step beyond binary detection.

\clearpage
\bibliographystyle{IEEEbib}
\bibliography{strings,refs}

\end{document}